\documentclass[aps,pra,showkeys,preprint,amsmath,amssymb]{revtex4}
\usepackage{graphicx,epstopdf}

\begin{document}

\title{Phase transitions and spectral singularities in a class of one-dimensional  parity-time-symmetric  complex potentials}

\author{Jinlin Fan}
\affiliation{College of Physics and Electronic Engineering \& Key Laboratory of Laser Technology and Optoelectronic Functional Materials of Hainan Province, Hainan Normal University, Haikou 571158, China}

\author{Feilong Wang}
\affiliation{College of Physics and Electronic Engineering \& Key Laboratory of Laser Technology and Optoelectronic Functional Materials of Hainan Province, Hainan Normal University, Haikou 571158, China}

\author{Ruolin Chai}
\affiliation{Center for Theoretical Physics \& School of Physics and Optoelectronic Engineering, Hainan University, Haikou 570228, China}

\author{Zhibin Zhao}
\email[Email: ]{zhaozhibin@hainnu.edu.cn}
\affiliation{College of Physics and Electronic Engineering \& Key Laboratory of Laser Technology and Optoelectronic Functional Materials of Hainan Province, Hainan Normal University, Haikou 571158, China}

\author{Qiongtao Xie}
\email[Corresponding author: ]{xieqiongtao@hainnu.edu.cn; xieqiongtao@126.com}
\affiliation{College of Physics and Electronic Engineering \& Key Laboratory of Laser Technology and Optoelectronic Functional Materials of Hainan Province, Hainan Normal University, Haikou 571158, China}

\begin{abstract}
We investigate a two-parametric family of one-dimensional non-Hermitian complex potentials with parity-time ($\mathcal{PT}$) symmetry. We find that there exist two distinct types of phase transitions, from an unbroken phase (characterized by a real spectrum) to a broken phase (where the spectrum becomes complex). The first type involves the emergence of a pair of complex eigenvalues bifurcating from the continuous spectrum. The second type is associated with the collision of such pairs at the bottom of the continuous spectrum. The first transition type is closely related to spectral singularities (SSs), at which point the transmission and reflection coefficients are divergent simultaneously. The second is associated with the emergence of bound states. In particular, under specific parameter conditions, we construct an exact bound state solution. By systematically exploring the parameter space, we establish a universal relationship governing the number of SSs in these potentials. These findings provide a fundamental theoretical framework for manipulating wave scattering in non-Hermitian systems, offering promising implications for designing advanced optical and quantum devices.

 \keywords{non-Hermitian complex potentials, phase transitions, spectral
singularities}

\end{abstract}
\maketitle

\section{Introduction}
Non-Hermitian systems with parity-time ($\mathcal{PT}$) symmetry have garnered significant interest over the past few decades due to their unique phenomena absent in Hermitian systems \cite{Kunst2021,Zezyulin2016}. $\mathcal{PT}$-symmetric systems, a subset of pseudo-Hermitian systems~\cite{Mostafazadeh2002}, exhibit a phase transition from an unbroken phase (real spectrum) to a broken phase (complex spectrum) as parameters vary \cite{Bender1998,Bender2007}. In general, this transition is governed by two distinct mechanisms: (i) exceptional points (EPs) in the discrete spectrum~\cite{Li2023} and (ii) spectral singularities (SSs) in the continuous spectrum~\cite{Mostafazadeh2009}. EPs arise from the coalescence of discrete eigenvalues and their corresponding eigenstates~\cite{Minganti2019}, while SSs are characterized by divergences in the reflection and transmission coefficients of scattering states~\cite{Mostafazadeh2012,Chong2011,Longhi2009}. Many interesting phenomena due to EPs and SSs are predicted, such as non-Hermitian skin effect \cite{Zhao2025,Yao2018,Yao2018b,Yokomizo2019}, non-Hermitian edge burst effect \cite{Zhu2024,Xiao2024}, non-Hermitian bulk-boundary correspondence \cite{Ao2020,Helbig2020,Kunst2018}, asymmetric mode conversion \cite{Ghosh2016}, nonreciprocal  transmission \cite{Peng2014}, unidirectional reflectionlessness~\cite{Yang2016,Feng2013},  coherent perfect absorption~\cite{Sun2014}, loss-induced transparency~\cite{Zhang2018}, chirality inversion \cite{Ni2024}, and  chiral mode switching~\cite{Heiss2016}. Parts of them have been verified in diverse systems, including optical structures~\cite{Guo2009,El-Ganainy2007,Benisty2011,Ruter2010}, acoustic setups~\cite{Fleury2015}, electrical circuits~\cite{Liu2021},  single-spin systems~\cite{Wu2019},  magneto-optical systems~\cite{Ruan2025}, and atomic systems \cite{Zhang2016}.

Beyond $\mathcal{PT}$-symmetric non-Hermitian systems, other classes retaining purely real spectra have been proposed, despite lacking PT symmetry. For example, by employing  supersymmetry method~\cite{Cannata1998, Miri2013}, operator symmetry method~\cite{Nixon2016, Yang2017, Bagchi2020}, and  soliton theory~\cite{Kawabata2020}, various classes of  non-$\mathcal{PT}$-symmetric complex potentials were constructed. Phase transitions in these systems can similarly be induced by parameter tuning. For specific one-dimensional (1D)  localized  complex potentials, such transitions involve the emergence of a pair of complex eigenvalues bifurcating from the continuous spectrum \cite{Yang2017}. This phenomenon represents the first instance of a phase transition occurring without EPs. Subsequent research revealed that this transition relates to the splitting of self-dual SSs \cite{Konotop2019}. In addition, a universal form of localized complex potentials with  multiple SSs has been presented ~\cite{Zezyulin2020}.  Interestingly, exact analytical solutions for SSs in such potentials have been found. Furthermore, the extension into the nonlinear situations has been studied \cite{Mostafazadeh2013a,Mostafazadeh2013b,Agarwal2014}.

Despite extensive study on SSs, a systematic study of the relationship between SSs and phase transitions in localized complex potentials remains lacking. Furthermore, existing work has not established an explicit quantitative link between the number of SSs and the potential's parameters. Resolving this critical issue would provide a fundamental theoretical foundation for controlling scattering properties in non-Hermitian systems.
This work investigates a class of one-dimensional (1D) non-Hermitian $\mathcal{PT}$-symmetric potentials featuring two tunable parameters. We identify two distinct types of phase transitions from an unbroken (real spectrum) to a broken phase (complex spectrum). The first type involves the emergence of complex eigenvalue pairs bifurcating from the continuous spectrum. The second type occurs when such complex eigenvalue pairs collide at the spectral edge. The first transition is closely linked to SSs, characterized by diverging transmission and reflection coefficients. The second transition correlates with bound state formation. Furthermore, we analytically present an exact bound state solution under specific parameter conditions. Through systematic exploration of the parameter space, we uncover a universal law governing the number of SSs possible within this potential class. These findings provide fundamental insights for manipulating scattering properties in engineered non-Hermitian systems.

\section{Theoretical models for a class of 1D non-Hermitian complex potentials}
We consider the Schr\"{o}dinger equation ($2 m=\hbar=1$)
\begin{equation}
-\frac{d^{2}}{d x^{2}} \psi(x)+V(x) \psi(x)=E \psi(x),\label{schE}
\end{equation}
where the potential function $V(x)$ takes the form~\cite{Nixon2016, Yang2017, Bagchi2020, Konotop2019, Zezyulin2020}
\begin{equation}
\begin{aligned}
V(x) & =V_r(x)+i V_i(x) \\
& =-W(x)^2-2 g W(x)-i W^{\prime}(x).
\end{aligned}
\end{equation}
Here, $W(x)$ denotes an arbitrary localized real function satisfying the boundary condition $\lim _{x \rightarrow \pm \infty} W(x)=0$, and $g$ represents a free parameter.
The functions ${V}_{r}({x})$ and ${V}_{i}({x})$ represent the real and imaginary components of the potential, respectively, while $W^{\prime}(x)$ indicates the first spatial derivative of $W(x)$.
This potential was proposed   through supersymmetry method~\cite{Cannata1998, Miri2013} and operator symmetry method~\cite{Nixon2016, Yang2017, Bagchi2020},  and has attracted considerable research interest.
This class of potentials exhibits entirely real spectra and phase transition characteristics. Notably, for any given real function $W(x)$, one can construct a corresponding complex potential.
Research has demonstrated that for specific choices of $W(x)$, tuning the parameter $g$ can induce a transition of the spectrum from entirely real to partially complex~\cite{Yang2017,Zezyulin2020}.

In the present study, we focus on the hyperbolic secant profile $W(x)=A \operatorname{sech}(x)$ ($A \in \mathbb{R}$), which generates the following complex potential structure
\begin{equation}
\begin{aligned}
V(x) & =-A^2 \operatorname{sech}^2(x)-2 g A \operatorname{sech}(x)+i A \operatorname{sech}(x) \tanh (x).
\end{aligned}\label{potV}
\end{equation}
This  potential  possesses $\mathcal{PT}$-symmetry and incorporates two tunable parameters, $g$ and $A$. We find that by tuning the parameter $g$ with the parameter $A$ fixed, there are phase transitions where a pair of complex eigenvalues bifurcate  from the continuous spectrum. We further reveal that such phase transitions are associated with the  SSs \cite{Ahmed2001a,Ahmed2001,Jones2007}.

\subsection{Phase transitions and SSs}
\begin{figure}[htbp]
\centering
\includegraphics[width=0.8\linewidth]{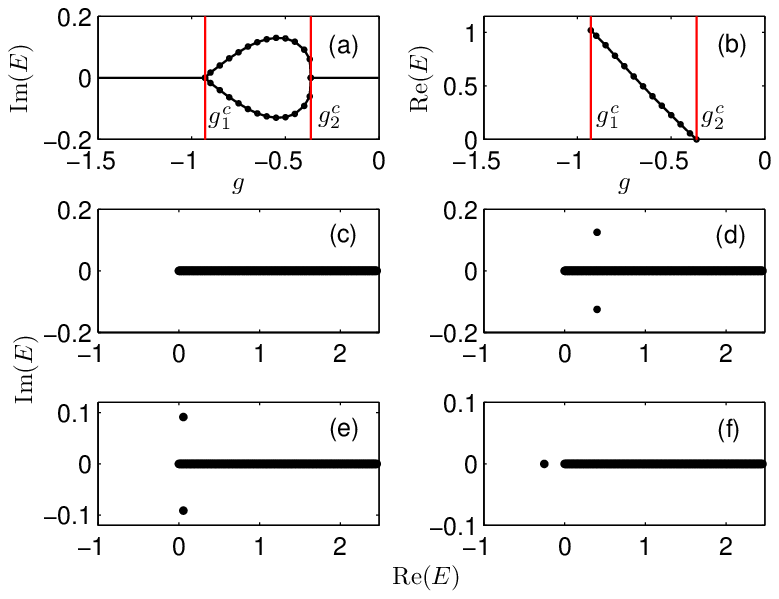}
\caption{(a) Imaginary part of the energy eigenvalues, $\operatorname{Im}(E)$, as a function of the  parameter $g$ for $A=3/2$. (b) Real parts of the complex eigenvalues between the two phase transitions points $g=g_{c,1}=-0.928$ and $g=g_{c,2}=-0.365$,  indicated by the two vertical lines in both (a) and (b).
(c)-(f) Energy spectra in the complex plane for  $A=3/2$  at specific values of $g$: (c) $g=-1$, (d) $g=-3/5$, (e) $g=-2/5$, and (f) $g=-1/2\sqrt{3}=-0.289$.  }
\label{fig:fig1}
\end{figure}

 We fix the potential parameter  $A=1$, vary another  potential parameter $g$, and investigate its effect  on the energy spectrum of the complex potential.  Figures~\ref{fig:fig1}(a) and (b), we show  the dependence of the real and imaginary parts of the energy eigenvalues, $\mathrm{Re}(E)$ and $\mathrm{Im}(E)$, on   $g$,  respectively. Figure ~\ref{fig:fig1}(a) reveals two phase transitions.  The first one appears  at $g_{c,1}=-0.928$, where a pair of complex-conjugate eigenvalues bifurcates from the continuum spectrum at energy  $E_{c,1}=1.038$ (corresponding to the wave numbers $k_{c,1}=\pm 1.019$). The second one appears at $g_{c,2}=-0.365$, where a pair of complex-conjugate eigenvalues meet at the bottom of the continuous spectrum with  $E_{c,2}=0(k_{c,2}=0)$ (see also Fig.~\ref{fig:fig1}(a)). These transitions are further illustrated in Figs.~\ref{fig:fig1}(c)-(f), which display the spectrum for specific values of $g$. For $g=-1<g_{c,1}$, only the real continuum spectrum exists ( Fig.~\ref{fig:fig1}(c)). For $g=-3/5$ and $g=-2/5$ in the range of $g_{c,1}<g<g_{c,2}$, a complex-conjugate energy pair splits off from the continuum, indicating the emergence of the $\mathcal{PT}$-broken phase (Figs.~\ref{fig:fig1}(d) and (e)).  For $g=-1/2\sqrt{3}=-0.289>g_{c,2}$, a discrete bound state emerge( Fig.~\ref{fig:fig1}(f)). Notably, this bound state admits an exact solution under the condition
\begin{equation}
A = \frac{1}{1 - 4g^2}.
\end{equation}
The corresponding wave function is
\begin{equation}
\psi(x)=e^{(A-1)x}\left(e^x-i\right)^{-A}\left(e^x+i\right)^{1-A}\left(1-\frac{\left(2 g+i\right) }{1+2 i g}e^x\right),
\end{equation}
with energy eigenvalue
\begin{equation}
E=-\frac{16 g^4}{\left(1-4 g^2\right)^2}.
\end{equation}
This exact solution is valid for $g^2<1/4$ (implying $A>1$). It can be verified by substituting it in Eq.~(\ref{schE}), and also provides an efficient check for the numerical results.  For the case of $A=3/2$, we have $g=\pm 1/2\sqrt{3}=\pm 0.289$ and $E=-1/4$,  matching the numerical result in Fig.~\ref{fig:fig1}(f).

In the following, we show that the first phase transition is associated with SS. We employ the transfer matrix method for numerical computation~\cite{Mostafazadeh2012}. The general solution of Eq.~(\ref{schE}) has the asymptotic behavior
\begin{eqnarray}
\psi(x) &\sim& a e^{i k x}+b e^{-i k x}, \quad x \rightarrow-\infty,\\
\psi(x) &\sim& c e^{i k x}+d e^{-i k x}, \quad x \rightarrow \infty,
\end{eqnarray}
where $k$ is the wave number with the incident energy $E=k^2$. The transfer matrix $M\left(g, k\right)$ relates the wave function amplitudes on the left $\binom{a}{b}$ to those on the right $\binom{c}{d}$
\begin{equation}
\binom{c}{d}=M\left(g, k\right)\binom{a}{b}. 
\end{equation}
Here, $M(g, k)$ is a $2 \times 2$ matrix:
\begin{equation}
M\left(g, k\right)=\left(\begin{array}{ll}
M_{11} & M_{12} \\
M_{21} & M_{22}
\end{array}\right).
\end{equation}
The transfer matrix $M\left(g, k\right)$ determines the transmission and reflection coefficients for both left and right incidence~\cite{Mostafazadeh2012}
\begin{equation}
T=\left|\frac{1}{M_{22}}\right|^{2}, \quad R^{l}=\left|\frac{M_{21}}{M_{22}}\right|^{2}, \quad R^{r}=\left|\frac{M_{12}}{M_{22}}\right|^{2}.
\end{equation}
A SS occurs when $M_{22}=0$ at critical values $g=g_*$ and $k=k_{2,*}$
\begin{equation}
M_{22}\left(g_*, k_{2,*}\right)=0.
\end{equation}
which causes these transmission and reflection coefficients  to diverge simultaneously.  Similarly, if $M_{11}=0$ at  $g=g_*$ and $k=k_1^*$, it defines a time-reversed SS (TRSS). In the $\mathcal{PT}$-symmetric system studied here,  $M_{11}=M_{22^*}$ \cite{Konotop2019},  so the SS and  TRSS coincide at the same  $g_*$ and $k_*$, implying  $M_{11}\left(g_*, k_*\right)=M_{22}\left(g_*, k_*\right)=0$. Thus we analyze $M_{22}$ to study SSs.
\begin{figure}[t]
\centering
\includegraphics[width=0.6\linewidth]{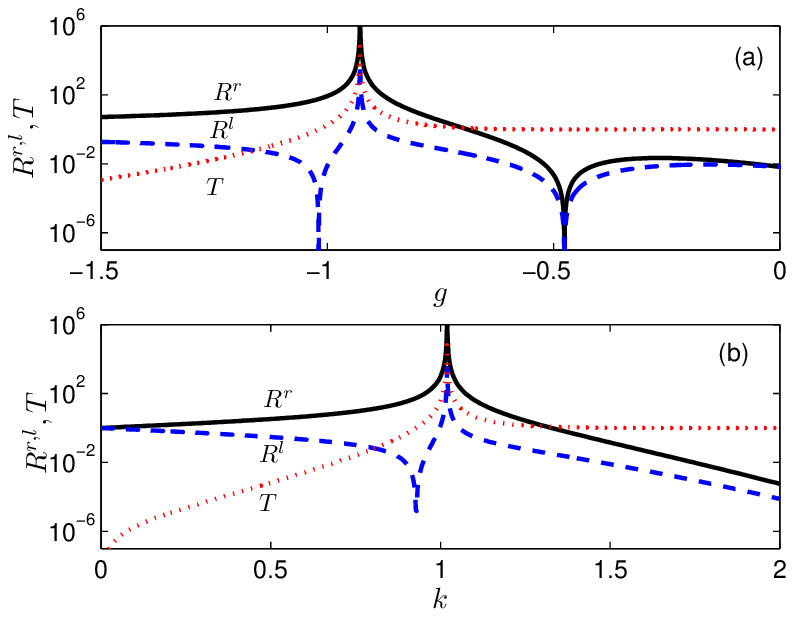}
\caption{ Reflection coefficients $R^{r}(k)$ (solid lines) and $R^{l}(k)$ (dashed lines),  and transmission coefficient $T(k)$ (dotted lines) as functions of (a) $g$ ( $A=3/2$ and $k=1.019$), and (b) $k$ ($A=3/2$ and $g=-0.929$). }
\label{fig:fig2}
\end{figure}

Generally, by systematically tuning the relevant parameters, $A$ and $g$, SSs can be identified through the reflection and transmission coefficients. Numerical results for  $A=3/2$ are presented in Fig.~\ref{fig:fig2}.   Figure ~\ref{fig:fig2}(a) shows the reflection and transmission coefficients,   $R^{r,l}(k)$ and  $T(k)$,  as functions of $g$ at the fixed wave number $k=k_{c,1}=1.019$.  Figure ~\ref{fig:fig2} shows the same coefficients as functions of  $k$ with  $g=g_{c,1}=-0.929$. We observe that  $R^{r,l}(k)$ and  $T(k)$ diverge to infinity at  $g=g_{c,1}=-0.929$ and $k=k_{c,1}=1.019$, signifying the emergence of the first SS. Therefore, we have  $g_*=g_{c,1}$ and $k_*=k_{c,1}$.   This identifies $g_*=g_{c,1}=-0.929$ and $k_*=k_{c,1}=1.019$.  Therefore, the emergence of an SS correlates with a phase transition originating from the continuous spectrum. This phase transition provides a fundamental characterization of SSs.

\subsection{Universal dependence of the number of SSs on the parameter $A$}
SSs represent distinct features in non-Hermitian systems.  As discussed in the previous section using the transfer matrix $M(g,k)$
In the previous section, SSs correspond to real zeros of the matrix element $M_{22}(g,k)$ at critical values  $g=g_*$ and $k=k_*$, and mark phase boundaries where the transition from the continuous spectrum occurs. In particular, we show that SSs are closely related to the  phase transition from the continuous spectrum. From the phase transition points,  we obtain  the relevant parameters $g=g_*$ and $k=k_*$.

\begin{figure}[htbp]
\centering
\includegraphics[width=0.6\linewidth]{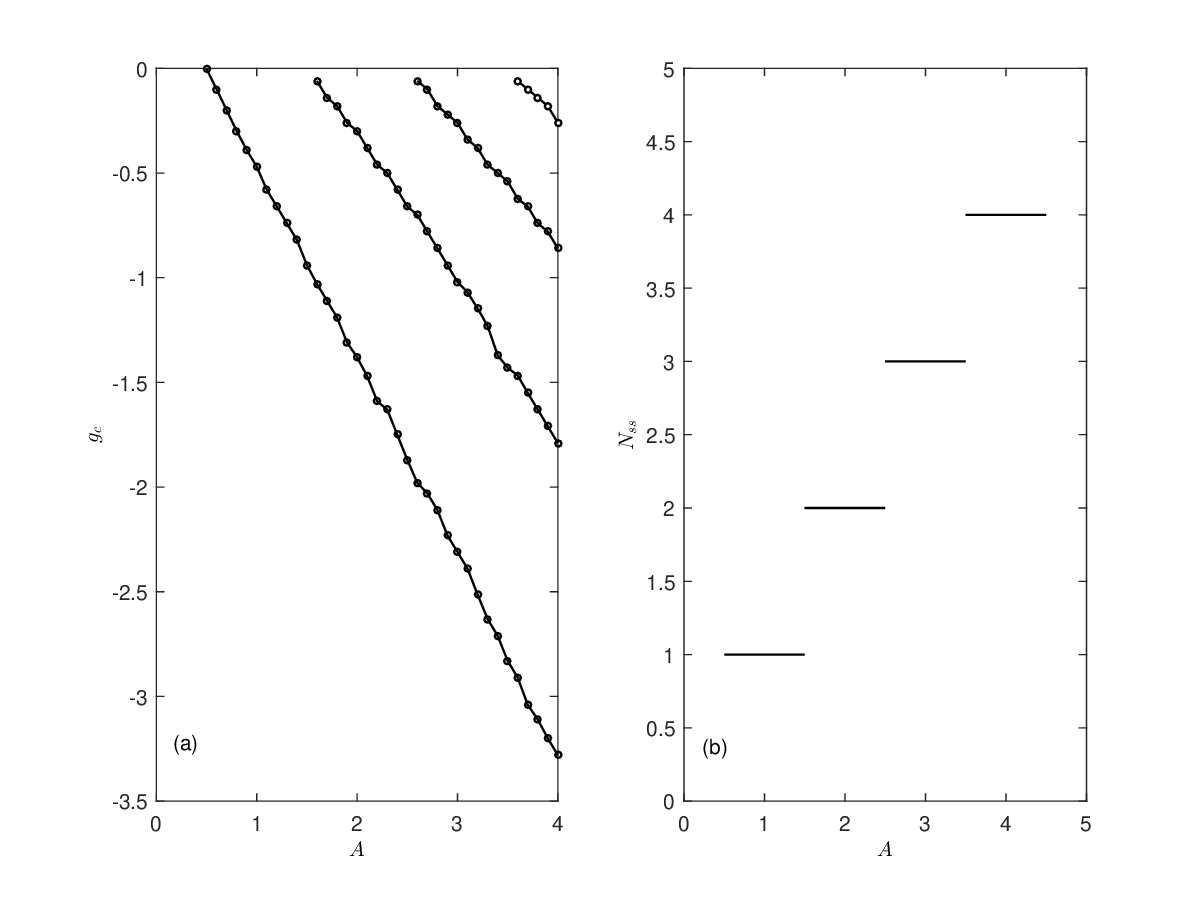}
\caption{(a) Critical values  $g_c$  associated with the phase transition from the continuous spectrum as a function of $A$. (b) Number of SSs, $N_{ss}$,  as a function of  the parameter $A$. }
\label{fig:fig3}
\end{figure}

While Ref.\cite{Zezyulin2020} established complex potentials supporting multiple SSs, this work demonstrates the controlled emergence of SSs via parameter $A$.  For A=3/2 (Fig.\ref{fig:fig2}), a single SS exists.  Numerical analysis reveals that increasing $A$ generates additional SSs. Figure~\ref{fig:fig3}(a) plots the phase transition thresholds $g_c$ versus $A$. 
Ii is observed that   no  SS  for $A<1/2$, one SS  $1/2\leq A<3/2$ ( single $g_c$), and two SSs for  $3/2\leq A <5/2$ (two  $g_c$ values).  The universal relationship between the SS count $N_{ss}$ and $A$ is
\begin{equation}
 N_{ss}=n+1 \mathrm{when} n+\frac{1}{2} \leq A < n+1+\frac{1}{2}\emph{}, \quad n=0,1,2,3,4,\ldots.
\end{equation}
as quantified in  Fig.~\ref{fig:fig3}(b).  This discrete scaling of SS density demonstrates precise control over potential features $V(x)$ via $A$. Targeting specific $A$ values enables deterministic positioning of multiple SSs, offering new pathways for engineering non-Hermitian phenomena.

Our results reveal that the number of SSs exhibits a characteristic regular distribution within specific parameter regimes. This finding highlights the robust control of parameter $A$ over the potential function $V({x})$. Through judicious selection of $A$ values, multiple SSs can be generated at designated positions, thereby opening new possibilities for manipulating SSs features in non-Hermitian systems.

\section{Conclusions}
In this work, we investigate a class of 1D non-Hermitian complex potentials with tunable parameters.  Through parameter modulation, the system can undergo phase transitions characterized by the emergence of complex-conjugate eigenvalue pairs.

 We identify two distinct types of phase transitions from an unbroken (real spectrum) to a broken phase (complex spectrum). The first type involves the emergence of complex eigenvalue pairs bifurcating from the continuous spectrum. The second type occurs when such complex eigenvalue pairs collide at the spectral edge. The first transition is closely linked to SSs, characterized by diverging transmission and reflection coefficients. The second transition correlates with bound state formation. Furthermore, our study reveals that  SSs appear in the system, manifested by the simultaneous divergence of both reflectance and transmittance, indicating the occurrence of a phase transition. Through systematic investigation of the  mechanism between SSs and phase transitions,  phase transitions  in the continuous spectrum are closely associated with  SSs. Remarkably, the emergence of SSs is invariably accompanied by phase transition phenomena in the continuous spectrum, unveiling profound connections between these fundamental characteristics of non-Hermitian systems.

By precisely tuning the relevant parameters, we discovered multiple SSs across different parameter intervals and established a quantitative relationship between the number of SSs and certain potential parameters. This finding not only advances the understanding of non-Hermitian scattering anomalies, but also provides a tunable theoretical framework for controlling SSs in quantum optics, $\mathcal{PT}$-symmetric lasers, and metamaterials. However, the current study has not fully elucidated the underlying physical mechanisms of this phenomenon. Future work must address two critical issues: First, whether the observed correlation between SSs and the parameter $A$ holds for a broader class of complex potentials; second, whether a universal theoretical framework can be established to uniformly describe the quantitative relationship between SSs and tuning parameters across different potential functions. Addressing these questions will significantly deepen our fundamental understanding of anomalous scattering behaviors in non-Hermitian quantum systems.

\end{document}